\documentclass[twocolumn]{svjour3} 

\usepackage{mathptmx}      
\smartqed  

\usepackage{amsmath}
\usepackage{graphicx}
\usepackage{bm}
\usepackage{stmaryrd}

\usepackage{epstopdf}

\ifdefined\rd
\else
\newcommand{\rd}{\,\mathrm{d}}
\fi

\newcommand{\pfrac}[2]{\frac{\partial #1}{\partial #2}}

\newcommand{\V}[1]{\bm{#1}}

\ifdefined\sgn
\else
\newcommand{\sgn}[1]{\mbox{sgn$\left(\V{#1}\right)$}}
\fi

\newcommand{\wcal}{\mathcal{W}}

\graphicspath{{Images/},{Figures/},{FiguresMatlab/}}
\usepackage{graphicx}

\usepackage{amsfonts, amssymb, amsbsy, amsmath}
\usepackage{pstricks,psfrag}

\renewcommand{\[}{\begin{equation}}
\renewcommand{\]}{\end{equation}}

\newcommand{\ie}{i.e., }
\newcommand{\sumj}{\sum_{j=1,j\not=i}^N\!\!}
\newcommand{\sumk}{\sum_{k=N+1}^{N+K}\!\!}
\newcommand{\sumi}{\sum_{i=1}^N}
\newcommand{\wcalij}{\int_0^1  \wcal (\V{r} - \V{r}_i + s \V{r}_{ij}) \rd s}
\newcommand{\wcalik}{\int_0^1  \wcal (\V{r} - \V{r}_i + s \V{a}_{ik})  \rd s}
\newcommand{\rg}{$-\int_z^\infty g_z \rho dr_z$}
\newcommand{\si}{$\sigma_{zz}+\int_z^\infty t_z dr_z$}

\journalname{}
\begin{document}
\title{From discrete particles to continuum fields near a boundary}
\author{Thomas Weinhart$^{1,2,\dagger}$ \and Anthony R. Thornton${}^{1,2}$ \and Stefan Luding$^{1}$ \and Onno Bokhove$^{2}$}
\institute{
${}^{1}$ Multiscale Mechanics, Dept. of Mechanical Engineering\\
${}^{2}$ Num. Analysis and Comp. Mechanics, Dept. of Applied Mathematics\\
${}^{1,2}$ Univ. of Twente, P.O.\,Box 217, 7500\,AE Enschede, The Netherlands\\ 
${}^{\dagger}$ \email{t.weinhart@utwente.nl}, Tel.: +31 53 489 3301
}
\maketitle

\begin{abstract}
An expression for the stress tensor near an external boundary of a discrete mechanical system is derived explicitly in terms of the constituents' degrees of freedom and interaction forces. Starting point is the exact and general coarse graining formulation presented by Goldhirsch in [I. Goldhirsch, Gran. Mat., 12(3):239-252, 2010], which is consistent with the continuum equations everywhere but does not account for boundaries. Our extension accounts for the boundary interaction forces in a self-consistent way and thus allows the construction of continuous stress fields that obey the macroscopic conservation laws even within one coarse-graining width of the boundary. 

The resolution and shape of the coarse-graining function used in the formulation can be chosen freely, such that both microscopic and macroscopic effects can be studied. The method does not require temporal averaging and thus can be used to investigate time-dependent flows as well as static and steady situations. Finally, the fore-mentioned continuous field can be used to define \lq fuzzy\rq\ (highly rough) boundaries.  Two discrete particle method  (DPM) simulations are presented in which the novel boundary treatment is exemplified, including a chute flow over a base with roughness greater than a particle diameter.
\end{abstract}


\section{Introduction}\label{sec:int}

The main topic of this paper is the issue of coarse-graining, near a boundary. We consider the bulk method described by Isaac Goldhirsch \cite{Goldhirsch2010}, and extend it to account for boundary forces due to the presence of a wall or base. The Goldhirsch special edition of Granular Matter is an appropriate place to present some of the ideas that we have developed in this area.

Continuum fields often need to be constructed from from discrete particle data. In molecular dynamics \cite{FrenkelSmit1996} and granular systems \cite{CundallStrack1979,Luding2008b}, these discrete data are the positions, velocities and forces of each atom or particle. In contrast, in the case of smooth particle hydrodynamics \cite{Monaghan2005}, the continuum system itself is approximated by a discrete set of fluid parcels. In all these methods, a crucially important issue is how to compute the continuum fields in the most appropriate way.
Several techniques have been developed to calculate the continuum fields, see \cite{LudingAllonso-Marroquin2011} and references therein.
Particularly the stress tensor is of interest: the techniques include the Irvin-Kirkwood's approach \cite{IrvingKirkwood1950} or the me\-thod of planes \cite{ToddEvansDaivis1995}. Here, we use the coarse-graining approach (CG) as first described in \cite{Babic1997}. 

The CG method \cite{Babic1997,Goldhirsch2010} has several advantages over other methods, including: \emph{i)} the fields automatically satisfy the conservation equations of continuum mechanics; \emph{ii)} it is not assumed that the particles are rigid or spherical, and \emph{iii)} the results are valid for single particles (no averaging over ensembles of particles is required). The only assumptions are: each particle pair has a single point of contact, the contact area can be replaced by a contact point, and collisions are not instantaneous. 

In Sect.\ \ref{sec:theory}, we use the derivation of \cite{Goldhirsch2010} to extend the CG method to account for the presence of a boundary. 
Explicit expressions for the resulting continuum fields are derived. 
In Sect.\ \ref{sec:extending}, an alternative stress definition is proposed extending the stress field into the boundary region.
In Sect.\ \ref{sec:results}, the approach is tested with two DPM simulations, and in Sect.\ \ref{sec:conc} we draw conclusions.

\section{Theory}\label{sec:theory}


\subsection{Assumptions and notation} \label{sec:forces}

We are interested in deriving macroscopic fields, such as density, velocity and the stress tensor from averages of microscopic variables such as the positions, velocities and for\-ces of the constituents. Averaging will be done such that the continuum fields, by construction, satisfy conservation laws.
Vectorial and tensorial components are denoted by Greek letters in order to distinguish them from the Latin particle-indices $i,j$. Bold vector notation will be used when appropriate. We will follow the derivation of \cite{Goldhirsch2010}, but extend it by introducing two types of particle: $N$ flowing particles $\{1,2,\dots,N\}$ and $K$ boundary particles $\{N+1,\dots,N+K\}$. 

Each particle $i$ has mass $m_i$, center of mass position $r_{i\alpha}$ and velocity $v_{i \alpha}$. 
The force $f_{i\alpha}$ acting on particle $i$ is a combination of the sum of the interaction force $f_{ij \alpha }$ with another particle $j$, the interaction force $f_{ik \alpha}$ with a boundary particle $k$, and a body force $ b_{i \alpha}$ (e.g., gravity),
\[ f_{i\alpha } = \sumj f_{ij \alpha }+\sumk f_{ik \alpha }+b_{i\alpha },\quad i\leq N.
\label{eq:force}\]
The interaction forces are binary and anti-symmetric such that action equals reaction, $f_{ij \alpha }=-f_{ji \alpha }$, $i,j\leq N$. 
We assume that each particle pair $(i,j)$, $i\leq N$, $j\leq N+K$ has, at most, a single contact point, $c_{ij \alpha}$, at which the contact forces act. 
The	positions of the boundary particles are fixed, as if they had infinite mass. 
The trajectories of the flowing particles are governed by Newton's second law and if tangential forces and torques are present, rotations follow from the angular form of Newton's law.

In the following sections, we commence from Ref. \cite{Goldhirsch2010} to derive definitions of the continuum fields. 
To be precise,  a \emph{body force density} is introduced to account for body forces, and to incorporate boundary effects an \emph{interaction force density} (IFD) is introduced. While the idea of an IFD is more generally applicable (e.g., for mixtures), it is employed here to account for the presence of a boundary.

\subsection{Coarse graining} \label{sec:cg}

From statistical mechanics, the microscopic mass density of the flow at a point $r_{\alpha}$ at time $t$ is defined by 
\begin{equation}
\rho^{\mbox{mic}} (\V{r},t) = \sumi m_i \delta \left(\V{r} -\V{r}_i(t) \right),
\end{equation} 
where $\delta(\V{r})$ is the Dirac delta function.
We use the following definition of the macroscopic density,
\begin{equation}\label{eq:sd:2}
\rho (\V{r},t) = \sumi m_i \wcal \left(\V{r} -\V{r}_i(t)\right),
\end{equation}
\ie we have replaced the Dirac delta function by an integrable \lq coarse-graining\rq\ function $\wcal$ whose integral over the domain is unity.

\subsection{Mass balance}\label{sec:mass}

The coarse-grained momentum density is defined by
\begin{equation}\label{eq:sd:5}
p_{\alpha}(\V{r},t) = \sumi m_i v_{i \alpha} \wcal (\V{r} - \V{r}_i).
\end{equation}
Hence, the macroscopic velocity field $V_{\alpha}(\V{r},t)$ is defined as the ratio of momentum and density fields, 
$V_{\alpha}(\V{r},t) = $ \linebreak[4]$p_{\alpha}(\V{r},t)/\rho(\V{r},t)$.
It is straightforward to confirm that $\rho_{\alpha}$ and $p_{\alpha}$  satisfy the continuity equation (c.f. \cite{Goldhirsch2010,Babic1997}),
\begin{equation}\label{eq:mass}
\pfrac{\rho}{t} + \pfrac{p_{\alpha}}{r_{\alpha}} = 0.
\end{equation}

\subsection{Momentum balance}\label{sec:momentum}

Subsequently, we will consider the momentum conservation equation with the aim of establishing the macroscopic stress field, $\sigma_{\alpha\beta}$. As we want to describe boundary stresses as well as internal stresses, the boundary interaction force density (IFD), $t_\alpha$, has been included, as well as the body force density, $b_{\alpha}$, which are not present in the original derivation, \cite{Goldhirsch2010}. The desired momentum balance equations take the form,
\begin{equation}\label{eq:momentum}
\pfrac{p_{\alpha}}{t} = - \pfrac{}{r_{\beta}}\left[\rho V_{\alpha} V_{\beta}\right] + \pfrac{\sigma_{\alpha \beta}}{r_{\beta}} + t_\alpha +  b_{\alpha}.
\end{equation}
%
To determine the stress it is required to compute the temporal derivative of \eqref{eq:sd:5},
\begin{eqnarray}\label{eq:sd:9}
\pfrac{p_{\alpha}}{t} 
&=& \sumi f_{i \alpha} \wcal(\V{r} -\V{r}_i)  + \sumi m_i v_{i \alpha} \pfrac{}{t}\wcal(\V{r} -\V{r}_i),
\end{eqnarray}
where $f_{i \alpha} = m_i \mbox{d} {v}_{i\alpha}/\mbox{d}t$ is the total force on particle $i$. Using \eqref{eq:force}, the first term in \eqref{eq:sd:9} can be expanded as
\begin{equation}\label{eq:sd:12}
A_{\alpha} \equiv \sumi \sumj f_{ij \alpha} \wcal_i + \sumi \sumk f_{ik \alpha} \wcal_i + \sumi  b_{i \alpha} \wcal_i,
\end{equation}
with the abbreviation $\wcal_i = \wcal(\V{r} - \V{r}_i)$.
The first term, which represents the bulk particle interactions, satisfies
\begin{eqnarray}\label{eq:sd:13}
\sumi \sumj f_{ij \alpha} \wcal_i &=& \sumi \sumj  f_{ji \alpha} \wcal_j 
\nonumber\\ &=&
- \sumi \sumj f_{ij \alpha} \wcal_j,
\end{eqnarray}
since $f_{ij\alpha} = -f_{ji \alpha }$ and because the dummy summation indices can be interchanged. It follows from \eqref{eq:sd:13} that 
\begin{eqnarray}\label{eq:sd:13b}
\sumi \sumj f_{ij \alpha} \wcal_i 
&=& \frac{1}{2} \sumi \sumj f_{ij \alpha} (\wcal_i-\wcal_j) \nonumber\\
&=& \sumi \sum_{j=i+1}^N f_{ij \alpha} \left(\wcal_i - \wcal_j\right).
\end{eqnarray}
Substituting \eqref{eq:sd:13b} into \eqref{eq:sd:12} yields
\begin{eqnarray}\label{eq:sd:14}
A_{\alpha} = \sumi \sum_{j=i+1}^N f_{ij \alpha} \left(\wcal_i - \wcal_j\right) +   \sumi \sumk f_{ik \alpha} \wcal_i   + b_{\alpha},
\end{eqnarray}
where $b_{\alpha}=\sum_i b_{i \alpha} \wcal_i$ is the body force density.

\smallskip
Next, $A_{\alpha}$ is rewritten using Leibnitz's rule to obtain a formula for the stress tensor. The following identity holds for any continuously differentiable coarse-graining function~$\wcal$
\begin{eqnarray}\label{eq:sd:15}
\wcal_j - \wcal_i &=& \int_0^1 \pfrac{}{s} \wcal (\V{r} - \V{r}_i + s \V{r}_{ij}) \rd s \nonumber\\ 
&=&r_{ij \beta} \pfrac{}{r_{\beta}} \wcalij,
\end{eqnarray}
where ${r}_{ij \alpha} =  r_{i \alpha} - r_{j \alpha}$ is the vector from $r_{j \alpha}$ to $r_{i\alpha}$. 
Substituting identities \eqref{eq:sd:15} into \eqref{eq:sd:14} yields
\begin{eqnarray}\label{eq:sd:16}
A_{\alpha} &=& 
-\pfrac{}{r_{\beta}}  \sumi \sum_{j=i+1}^N f_{ij \alpha} r_{ij \beta} \wcalij \nonumber\\
&&+\sumi \sumk f_{ik \alpha} \wcal_{i} + b_{\alpha}.
\end{eqnarray}
In Ref.\ \cite{Goldhirsch2010}, it is shown that the second term in \eqref{eq:sd:9} can be expressed as
\begin{equation}\label{eq:sd:16b}
\sumi m_i v_{i \alpha} \pfrac{}{t}\wcal_i = 
-\pfrac{}{r_{\beta}} \left[ \rho V_{\alpha} V_{\beta} + \sumi m_i v_{i \alpha}' v_{i \beta}' \wcal_i \right],
\end{equation}
where $v_{i \alpha}'$ is the fluctuation velocity of particle $i$, given by
\begin{equation}
v_{i \alpha}' (\V{r},t) =v_{i \alpha} (t) - V_{\alpha} (\V{r},t).
\end{equation}
Substituting \eqref{eq:sd:16} and \eqref{eq:sd:16b} into momentum balance \eqref{eq:momentum} yields
\begin{eqnarray}\label{eq:sd:20a}
\pfrac{\sigma_{\alpha \beta}}{r_{\beta}} + \underline{\!t_\alpha\!}
&\!=\!&\pfrac{\sigma_{\alpha \beta}^k}{r_{\beta}}+\pfrac{\sigma_{\alpha \beta}^b}{r_{\beta}} + \underline{\sumi \sumk f_{ik \alpha} \wcal_i}.
\end{eqnarray}
where the kinetic and bulk contact contributions to the stress tensor are defined as
\begin{subequations}\label{eq:stressparts}
\begin{eqnarray}
\sigma_{\alpha \beta}^k &=& -\sumi m_i v_{i \alpha}' v'_{i \beta} \wcal_i,\\
\sigma_{\alpha \beta}^b &=& -\!\sumi\! \sum_{j=i+1}^N\! f_{ij \alpha}\! r_{ij \beta}\! \wcalij. \label{eq:stressparts_b}
\end{eqnarray}
\end{subequations}
Here, the underlined terms in \eqref{eq:sd:20a} are not in the original derivation presented in Ref. \cite{Goldhirsch2010} and account for the presence of the boundary.

Expression \eqref{eq:sd:20a} can be satisfied by defining the last term on the right hand side as the IFD. This however has the disadvantage that the boundary IFD is located around the center of mass of the flowing particles. The more natural physical location of the boundary IFD would be at the interface between the flowing and boundary particles.

Therefore, we move the IFD to the contact points, $c_{ik \alpha}$, between flowing and boundary particles: similar to \eqref{eq:sd:15},
\begin{eqnarray}\label{eq:sd:15x}
\wcal_{ik} - \wcal_{i} &=&a_{ik \beta} \pfrac{}{r_{\beta}} \wcalik,
\end{eqnarray}
where $\wcal_{ik}=\wcal(\V{r}-\V{c}_{ik})$ and ${a}_{ik \alpha}={r}_{i\alpha}-{c}_{ik \alpha}$.
Substituting \eqref{eq:sd:15x} into the last term in \eqref{eq:sd:20a} we obtain
\begin{gather}
\sumi \sumk f_{ik \alpha} \wcal_i 
= \sumi \sumk f_{ik \alpha} \wcal_{ik} \nonumber\\
\phantom{W}-\pfrac{}{r_{\beta}} \left[\sumi \sumk f_{ik \alpha} a_{ik \beta} \wcalik \right].
\label{eq:sd:21}
\end{gather}
Thus, substituting \eqref{eq:sd:21} into \eqref{eq:sd:20a}, we define the stress by
\begin{subequations}\label{eq:stress}
\[\label{eq:stress_con}
\sigma_{\alpha \beta}=\sigma_{\alpha \beta}^k+\sigma_{\alpha \beta}^b+\sigma_{\alpha \beta}^w,\]
where the contribution to the stress from the contacts between flow and boundary particles is
\begin{eqnarray}
\sigma_{\alpha \beta}^w &=& - \sumi \sumk f_{ik \alpha} a_{ik \beta} \wcalik,
\end{eqnarray}
\end{subequations}
and the IFD is
\begin{equation} \label{eq:IFD}
t_\alpha = \sumi \sumk f_{ik \alpha} \wcal_{ik}.
\end{equation}

Equations \eqref{eq:stress} differ from the standard result of \cite{Goldhirsch2010} by an extra term, $\sigma_{\alpha\beta}^w$, that accounts for the additional stress created by the interaction of the boundary with the flow. The definition \eqref{eq:IFD} gives the boundary IFD applied by the flowing particles; \ie it has been constructed such that in the limit $w\to0$, the IFD acts at the contact points between boundary and flow. 

Note, that this framework is general and can be used to compute more than boundary IFDs. 
For example, one can obtain the drag between two different species of interacting particles by replacing the flowing and boundary particles with the particles of the two species in the definition of the continuum fields. By placing an IFD at the contact points, the IFDs of both species are exactly antisymmetric and thus disappear in the momentum continuity equation of the combined system. In mixture theory, e.g. \cite{Morland1992}, such interaction terms appear in the governing equations for the individual constituents and are called interaction body forces. These interaction body forces are an exact analog to the IFDs. Therefore, our approach can interpreted as treating the system as a mixture of boundary and flow particles and the IFD is the interaction body force between different species of particle.


Further, we note that the integral of the stress in \eqref{eq:stressparts} and \eqref{eq:stress} over the domain $\Omega$ satisfies the virial definition of mechanical stress, 
\begin{eqnarray}
\int_{\Omega} \sigma_{\alpha\beta} \rd \V{r} &=& 
- \sumi m_i v_{i \alpha}' v'_{i \beta}
\nonumber\\&&
- \sumi \sum_{j=i+1}^N f_{ij \alpha} r_{ij \beta}
- \sumi \sumk f_{ik \alpha} a_{ik \beta}.
\end{eqnarray}

\subsection{Extending the stress profile into a base or wall}\label{sec:extending}

In contrast to the previous subsection, an alternative stress definition is presented here, where the IFD and the stress are combined into a single tensor.
Similar to \eqref{eq:sd:15} and \eqref{eq:sd:15x}, the following identity holds,
\begin{eqnarray}\label{eq:sd:15inf}
-\wcal_i  
&=& \int_0^\infty \pfrac{}{s} \wcal (\V{r} - \V{r}_i + s \V{r}_{ik}) \rd s \nonumber\\ 
&=& r_{ik \beta} \pfrac{}{r_{\beta}} \int_0^\infty  \wcal (\V{r} - \V{r}_i + s \V{r}_{ik}) \rd s,
\end{eqnarray}
since the coarse-graining function $\wcal$ satisfies $\wcal(|\V{r}|\rightarrow \infty) =0$. 
Substituting \eqref{eq:sd:15inf} into \eqref{eq:sd:20a} we can obtain
an alternative solution with zero IFD, $t_\alpha'=0$, where the stress is given by
${\sigma_{\alpha \beta}}'=\sigma_{\alpha \beta}^k+\sigma_{\alpha \beta}^b+{\sigma_{\alpha \beta}^w}'$, with
\begin{eqnarray}\label{eq:stressinf}
{\sigma_{\alpha \beta}^w}' &=& - \sumi \sumk f_{ik \alpha} r_{ik \beta} \int_0^\infty  \wcal (\V{r} - \V{r}_i + s \V{r}_{ik})  \rd s.
\end{eqnarray}
This stress definition is not identical to the one in \eqref{eq:stress} and \eqref{eq:stressparts}. It eliminates the IFD term entirely and provides a natural extension of the stress into the boundary. 
However, the extended stress does contain contributions from both internal and external forces, and the 
spatial integral of the stress components has to be extended to infinity. 
In singular special cases this can lead to artificial results.
Another disadvantage of Eq.\ \eqref{eq:stressinf} is its difficult interpretation due to the long-ranging integral. One could see it as the stress inside a \lq virtual/fake\rq{}
wall-material on which the body-force is not acting (equivalent to foam with zero mass-density). However, this is far fetched and not realistic, so that we rather stick to the formulation in Sect. \ref{sec:momentum}.

It is also possible to extend the stress tensor to the boundary by other means, such as mirroring the stress at the boundary, or using a one-sided coarse-graining function. This is not discussed further since the first method requires a definition of the exact location of the boundary, while the second method can introduce a spatial shift in the stress field due to spatial inhomogeneities.

\section{Results}\label{sec:results}

\begin{figure*}[t]
\centering
\scalebox{1}{
\begin{psfrags}%
\psfragscanon%
%
\psfrag{s02}[t][t]{\fontsize{10}{15}\fontseries{m}\mathversion{normal}\fontshape{n}\selectfont \color[rgb]{0,0,0}\setlength{\tabcolsep}{0pt}\begin{tabular}{c}$x$\end{tabular}}%
\psfrag{s03}[b][b]{\fontsize{10}{15}\fontseries{m}\mathversion{normal}\fontshape{n}\selectfont \color[rgb]{0,0,0}\setlength{\tabcolsep}{0pt}\begin{tabular}{c}$z$\end{tabular}}%
\psfrag{s05}[b][b]{\fontsize{10}{15}\fontseries{m}\mathversion{normal}\fontshape{n}\selectfont \color[rgb]{0,0,0}\setlength{\tabcolsep}{0pt}\begin{tabular}{c}$|\V{t}|_2$\end{tabular}}%
\psfrag{s06}[][]{\fontsize{10}{15}\fontseries{m}\mathversion{normal}\fontshape{n}\selectfont \color[rgb]{0,0,0}\setlength{\tabcolsep}{0pt}\begin{tabular}{c} \end{tabular}}%
\psfrag{s07}[][]{\fontsize{10}{15}\fontseries{m}\mathversion{normal}\fontshape{n}\selectfont \color[rgb]{0,0,0}\setlength{\tabcolsep}{0pt}\begin{tabular}{c} \end{tabular}}%
%
\fontsize{10}{15}\fontseries{m}\mathversion{normal}%
\fontshape{n}\selectfont%
%
\psfrag{x01}[t][t]{0}%
\psfrag{x02}[t][t]{1}%
\psfrag{x03}[t][t]{2}%
\psfrag{x04}[t][t]{3}%
\psfrag{x05}[t][t]{4}%
\psfrag{x06}[t][t]{5}%
%
\psfrag{v01}[l][l]{0}%
\psfrag{v02}[l][l]{2}%
\psfrag{v03}[l][l]{4}%
\psfrag{v04}[l][l]{6}%
\psfrag{v05}[l][l]{8}%
\psfrag{v06}[l][l]{10}%
\psfrag{v07}[r][r]{-0.5}%
\psfrag{v08}[r][r]{0}%
\psfrag{v09}[r][r]{0.5}%
\psfrag{v10}[r][r]{1}%
\psfrag{v11}[r][r]{1.5}%
\psfrag{v12}[r][r]{2}%
%
\resizebox{8cm}{!}{\includegraphics{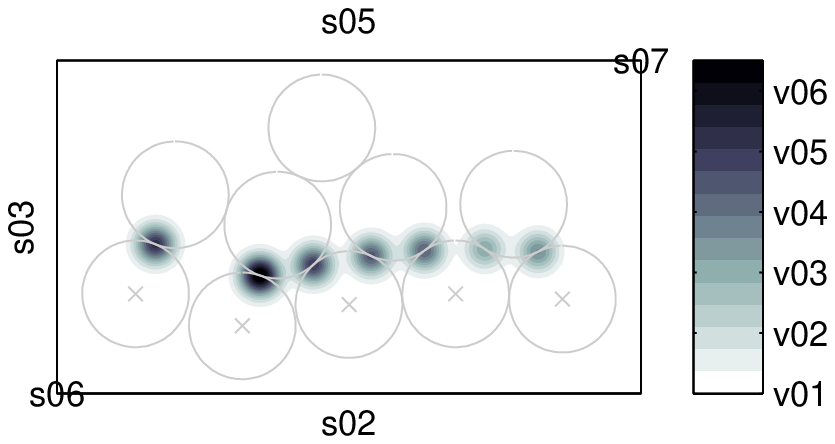}}%
\end{psfrags}%
%
}\quad
\scalebox{1}{
\begin{psfrags}%
\psfragscanon%
%
\psfrag{s02}[t][t]{\fontsize{10}{15}\fontseries{m}\mathversion{normal}\fontshape{n}\selectfont \color[rgb]{0,0,0}\setlength{\tabcolsep}{0pt}\begin{tabular}{c}$x$\end{tabular}}%
\psfrag{s03}[b][b]{\fontsize{10}{15}\fontseries{m}\mathversion{normal}\fontshape{n}\selectfont \color[rgb]{0,0,0}\setlength{\tabcolsep}{0pt}\begin{tabular}{c}$z$\end{tabular}}%
\psfrag{s05}[b][b]{\fontsize{10}{15}\fontseries{m}\mathversion{normal}\fontshape{n}\selectfont \color[rgb]{0,0,0}\setlength{\tabcolsep}{0pt}\begin{tabular}{c}$|\V{\sigma}|_2$\end{tabular}}%
\psfrag{s06}[][]{\fontsize{10}{15}\fontseries{m}\mathversion{normal}\fontshape{n}\selectfont \color[rgb]{0,0,0}\setlength{\tabcolsep}{0pt}\begin{tabular}{c} \end{tabular}}%
\psfrag{s07}[][]{\fontsize{10}{15}\fontseries{m}\mathversion{normal}\fontshape{n}\selectfont \color[rgb]{0,0,0}\setlength{\tabcolsep}{0pt}\begin{tabular}{c} \end{tabular}}%
%
\fontsize{10}{15}\fontseries{m}\mathversion{normal}%
\fontshape{n}\selectfont%
%
\psfrag{x01}[t][t]{0}%
\psfrag{x02}[t][t]{1}%
\psfrag{x03}[t][t]{2}%
\psfrag{x04}[t][t]{3}%
\psfrag{x05}[t][t]{4}%
\psfrag{x06}[t][t]{5}%
%
\psfrag{v01}[l][l]{0}%
\psfrag{v02}[l][l]{0.5}%
\psfrag{v03}[l][l]{1}%
\psfrag{v04}[l][l]{1.5}%
\psfrag{v05}[l][l]{2}%
\psfrag{v06}[l][l]{2.5}%
\psfrag{v07}[l][l]{3}%
\psfrag{v08}[r][r]{-0.5}%
\psfrag{v09}[r][r]{0}%
\psfrag{v10}[r][r]{0.5}%
\psfrag{v11}[r][r]{1}%
\psfrag{v12}[r][r]{1.5}%
\psfrag{v13}[r][r]{2}%
%
\resizebox{8cm}{!}{\includegraphics{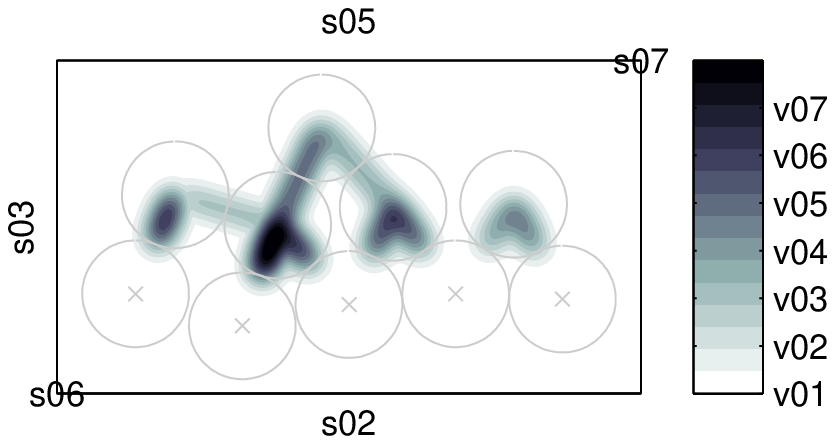}}%
\end{psfrags}%
%
}
\caption{Grey circles denote a two-dimensional configuration of free and fixed boundary particles, with gravity in $z$-direction. Fixed particles are marked with a cross in the center. Contour plots show the spatial distribution of the norms of the boundary IFD and the contact stress (magnitude of largest eigenvalue). A very small coarse graining width, $w=d/8$, is chosen to make the spatial averaging visible: the IFD centers around the contact point, while the stress is distributed along the contact lines. 
}
\label{fig:2D}
\end{figure*}

\begin{figure*}[t]
\begin{minipage}[b]{0.48\textwidth}
\centering
\scalebox{.3}{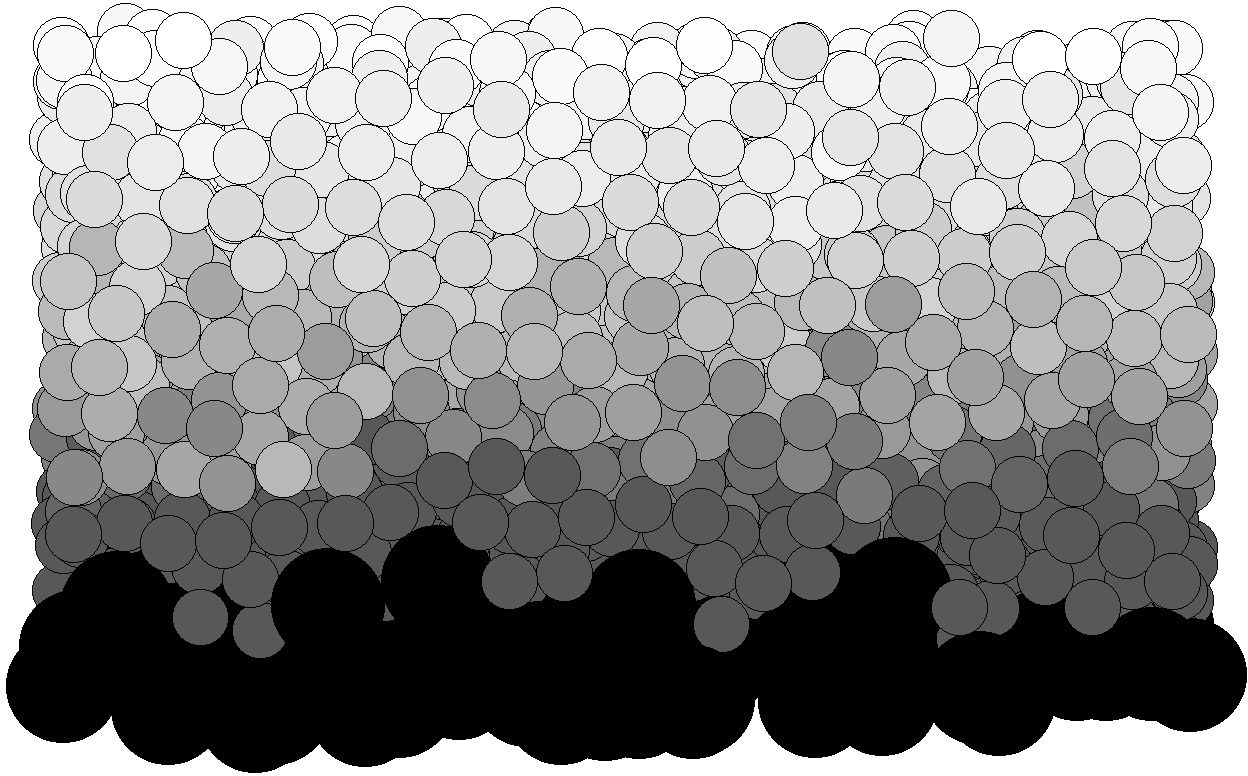}
\end{minipage}
\hfill
\begin{minipage}[b]{0.48\textwidth}
\centering
\scalebox{1.05}{
\begin{psfrags}%
\psfragscanon%
%
\psfrag{s01}[t][t]{\fontsize{10}{15}\fontseries{m}\mathversion{normal}\fontshape{n}\selectfont \color[rgb]{0,0,0}\setlength{\tabcolsep}{0pt}\begin{tabular}{c}z\end{tabular}}%
\psfrag{s05}[l][l]{\fontsize{10}{15}\fontseries{m}\mathversion{normal}\fontshape{n}\selectfont \color[rgb]{0,0,0}\rg}%
\psfrag{s06}[l][l]{\fontsize{10}{15}\fontseries{m}\mathversion{normal}\fontshape{n}\selectfont \color[rgb]{0,0,0}$\sigma_{zz}~~~~~$}%
\psfrag{s07}[l][l]{\fontsize{10}{15}\fontseries{m}\mathversion{normal}\fontshape{n}\selectfont \color[rgb]{0,0,0}\si}%
\psfrag{s08}[l][l]{\fontsize{10}{15}\fontseries{m}\mathversion{normal}\fontshape{n}\selectfont \color[rgb]{0,0,0}\rg}%
\psfrag{s10}[][]{\fontsize{10}{15}\fontseries{m}\mathversion{normal}\fontshape{n}\selectfont \color[rgb]{0,0,0}\setlength{\tabcolsep}{0pt}\begin{tabular}{c} \end{tabular}}%
\psfrag{s11}[][]{\fontsize{10}{15}\fontseries{m}\mathversion{normal}\fontshape{n}\selectfont \color[rgb]{0,0,0}\setlength{\tabcolsep}{0pt}\begin{tabular}{c} \end{tabular}}%
%
\fontsize{10}{15}\fontseries{m}\mathversion{normal}%
\fontshape{n}\selectfont%
%
\psfrag{x01}[t][t]{0}%
\psfrag{x02}[t][t]{2}%
\psfrag{x03}[t][t]{4}%
\psfrag{x04}[t][t]{6}%
\psfrag{x05}[t][t]{8}%
\psfrag{x06}[t][t]{10}%
%
\psfrag{v01}[r][r]{0}%
\psfrag{v02}[r][r]{2}%
\psfrag{v03}[r][r]{4}%
\psfrag{v04}[r][r]{6}%
\psfrag{v05}[r][r]{8}%
%
\resizebox{8cm}{!}{\includegraphics{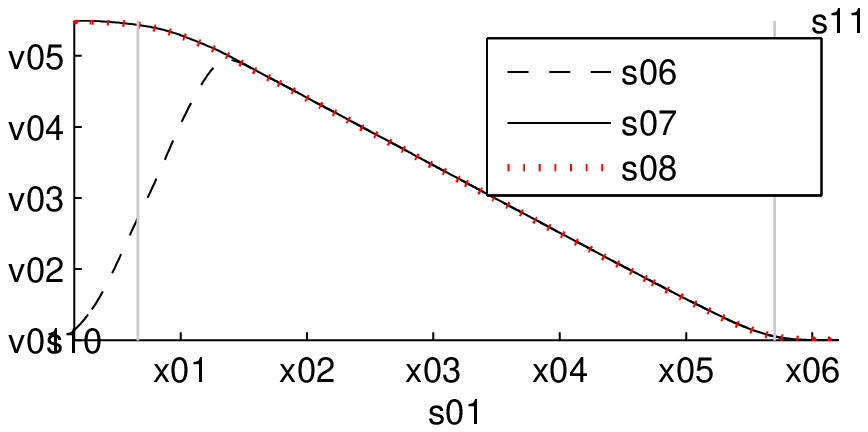}}%
\end{psfrags}%
%
}
\end{minipage}

\begin{minipage}[t]{0.48\textwidth}
\caption{Steady chute flow over a very rough frictional surface of inclination $\theta=26^\circ$ for $N=1000$ flowing particles. Gravity direction $\V{g}$ and coordinates $(x,y,z)$ as indicated. The domain is periodic in the $x$- and $y$-directions. Shade indicates speed; dark is slow and bright is fast.}
\label{fig:L2}
\end{minipage}
\hfill
\begin{minipage}[t]{0.48\textwidth}
\caption{Downward normal stress $\sigma_{zz}$ without (dashed) and with (solid) correction by the boundary IFD for $w=d/4$. The stress and IFD exactly match the weight of the flow above height $z$ (red dotted), as expected for steady flows. Grey lines indicate bed and surface location. 
}
\label{fig:L2Stress}
\end{minipage}
\end{figure*}

\subsection{Contact model} \label{sec:contact}

For illustrational purposes, we simulate a granular system with contact interaction forces. The statistical method, however, is based only on the assumptions in \S \ref{sec:theory} and therefore can be applied more generally. We use a viscoelastic force model with sliding friction as described in detail in \cite{SilbertErtasGrestHaleyLevinePlimpton2001,Luding2008b}. The parameters of the system are nondimensionalised such that the flow particle diameters are $d=1$,  their mass $m=1$,  and the magnitude of gravity $g=1$. The normal spring and damping constants are $k^n=2\cdot10^5$ and $\gamma^{\,n}=50$, respectively; thus, the collision time is $t_c=0.005\sqrt{d/g}$ and the coefficient of restitution is $\epsilon=0.88$. The tangential spring and damping constants are $k^{\,t}=2/7 k^n$ and $\gamma^{\,t}=\gamma^{\,n}$, such that the frequency of normal and tangential contact oscillation, and the normal and tangential dissipation are equal. The microscopic friction coefficient is set to $\mu^p=0.5$. We integrate the resulting force and torque relations in time using the Velocity-Verlet algorithm 
with a time step $\Delta t=t_c/50$.

We take the coarse-graining function to be a Gaussian of width, or variance, $w$.
Other coarse-graining functions are allowed, but the Gaussian has the advantage that it produces smooth fields and the required integrals can be performed exactly. 

\subsection{Quasi-static example in two dimensions}

In order to visualise definitions \eqref{eq:stress} and \eqref{eq:IFD}, we firstly consider a two-dimensional configuration consisting of five fixed boundary particles and five flowing bulk particles, with gravity in the $z$-direction, see Fig.\ \ref{fig:2D}. The flow is relaxed until the flowing particles are static; hence, the only contribution to the stress is due to the enduring contacts. To visualise the spatial distribution of the IFD and stress, the norms $|\V{t}|=\sqrt{t_\alpha^2}$ and $|\V{\sigma}|=\max_{|\V{x}|=1}|\V{\sigma}\V{x}|$ (the maximum absolute eigenvalue),
are displayed in Fig.\ \ref{fig:2D}. A very small coarse graining width, $w=d/8$, is chosen to make the spatial averaging visible: the IFD, Eq. \eqref{eq:IFD}, centers around the contact points between flowing and static particles, $r_{ik \alpha}$, while the stress, Eqs. \eqref{eq:stress} and \eqref{eq:stressparts}, is distributed along the contact lines, $\overline{r_{i \alpha}r_{j \alpha}}$ and $\overline{r_{i\alpha}c_{ik \alpha}}$.

\subsection{Three-dimensional steady chute flow}


Secondly, we consider a three-dimensional simulation of a steady uniform granular chute flow, see Fig.\ \ref{fig:L2} and Ref.\ \cite{WeinhartThorntonLudingBokhove2011}. The chute is periodic in the $x$- and $y$-directions and has dimensions $(x,y)\in[0,20]\times[0,10]$. The chute is inclined at $\theta=26^\circ$ and the bed consists of a disordered, irregular boundary created from fixed particles with size $d_{base}=2$. The chute contains $1000$ flowing particles, which are initially randomly distributed. The simulation is computed until a steady state is reached. A screen shot of the steady-state system is given in Fig.\ \ref{fig:L2}. 

Depth profiles for steady uniform flow are obtained by averaging with a coarse-graining width $w=d/4$ over $x\in[0,20]$, $y\in[0,10]$ and $t\in[2000,2100]$. The spatial averaging is done analytically, while we average in time with snapshots taken every $t_c/2$.

Note that the stress definitions \eqref{eq:stress} and \eqref{eq:stressinf} satisfy  
\begin{eqnarray}\label{eq:extendedstress}
\pfrac{\sigma_{\alpha \beta}'}{r_{\beta}} = \pfrac{\sigma_{\alpha \beta}}{r_{\beta}} + t_{\alpha}.
\end{eqnarray}
After averaging in $x$, $y$ and $t$ directions, this yields
\begin{equation}
\pfrac{\sigma_{\alpha z}'}{r_z}= \pfrac{\sigma_{\alpha z}}{r_z} + t_\alpha.
\end{equation}
Integrating over $(z,\infty)$, we obtain
\[ \sigma_{\alpha z}'=\sigma_{\alpha z}-\int_z^\infty t_\alpha dr_z.\label{eq:stressinf2}\]
Thus, setting $\alpha=z$ in \eqref{eq:stressinf2}, the extended stress component $\sigma_{zz}'$ can be obtained without computing the semi-infinite line integral.

 
The depth profile for the downward normal stress $\sigma_{zz}$ is shown in Fig.\ \ref{fig:L2Stress}. Since the rough boundary is not at a fixed height, the stress gradually decreases at the bottom due to the decreasing number of bulk particles near the base. Due to the coarse graining, the stress tensor has a gradient even in the case of a flat wall, but the gradient disappears as $w\to0$.
Using the extended stress definition, the bed and surface locations can be defined as the line where the downwards normal stress $\sigma_{zz}'$ vanishes and where it reaches its maximum value (to within 2\%), see Fig.\ \ref{fig:L2Stress}. 
Additionally, since the flow is steady and uniform, \eqref{eq:momentum} yields the so called lithostatic balance, $\sigma_{zz}'=-\int_z^\infty g_z\rho \,dr_z$,  
which is satisfied with good accuracy.
%


\section{Conclusions}\label{sec:conc}

We have derived explicit expressions for the stress tensor and the interaction force density (IFD) near an external boun\-da\-ry of a discrete mechanical system. These expressions were obtained by coarse-graining the microscopic equations and therefore exactly satisfy the governing balance laws of mass \eqref{eq:mass} and momentum \eqref{eq:momentum}. A boundary IFD was computed using the contact points between the flow
and the basal particles. 
%
Our results can be extended to other IFDs, for example, the drag between two different species of particles.
%
The power of our extension to Goldhirsch's method has been demonstrated by computing stress profiles for a chute flow over a fuzzy boundary. It avoids the problems inherent in other methods and gives the expected linear lithostatic profile all the way to the base.

The present formulation for boundary interaction forces allows us to draw the analogy to electrostatics, where the divergence of the electric field (analogous to the divergence of stress) is compensated by a charge-density source like our 
interaction force density \eqref{eq:IFD}.
The analogy  can also be made to mixture theory where, by treating the system as a mixture of boundary and flow particles, the IFD is then interpreted as the interaction body force between the two species.

\section{Acknowledgements}
The authors would like to thank 
the Institute for Mechanics, Process, and Control, Twente (IMPACT) and the NWO VICI grant 10828 for financial support, and Rem\-co Hartkamp and Dinant Krijgsman for fruitful discussions.


\end{document}